\documentclass[trackchanges, preprint2]{aastex701}

\usepackage{amsmath}
\usepackage{tikz}
\usepackage{CJK}
\usepackage{xcolor}
\usepackage{graphicx}
\usepackage{hyperref}
\usepackage{bm}
\usepackage{soul}
\usepackage{svg} 

\usepackage{CJK}
\defcitealias{goldreich_toward_1995}{GS95}
\defcitealias{boldyrev_2006}{B06}
\defcitealias{chandran_intermittency_2015}{CSM15}
\defcitealias{Mallet_2017}{MS17}
\defcitealias{iroshnikov_turbulence_1963}{IK}
\defcitealias{kolmogorov_local_1941}{K41}
\defcitealias{fox_solar_2016}{PSP}
\defcitealias{muller_solar_2020}{SolO}

\defcitealias{van1995python}{\texttt{Python}}
\defcitealias{2020SciPy-NMeth}{\texttt{SciPy}}
\defcitealias{mckinney2010data}{\texttt{Pandas}}
\defcitealias{Hunter2007Matplotlib}{\texttt{Matplotlib}}
\defcitealias{PYMC_2015}{\texttt{PyMC}}
\defcitealias{angelopoulos_space_2019}{\texttt{Pyspedas}}
\defcitealias{MHDTurbPy_Sioulas}{\texttt{MHDTurbPy}}

\begin{document}

\begin{CJK*}{UTF8}{gbsn}

\title{Generation and Expansion-Driven Growth of Switchbacks in the Outer Solar Corona and Solar Wind}

\author[0000-0002-1128-9685]{Nikos Sioulas}
\affiliation{Imperial College London, South Kensington Campus, London SW7 2AZ, UK} 
\email[show]{nsioulas@berkeley.edu}

\author[0000-0002-2381-3106]{Marco Velli}
\affiliation{Department of Earth, Planetary, and Space Sciences, University of California, Los Angeles, CA, USA}
\email{mvelli@ucla.edu}

\author[0000-0002-2582-7085]{Chen Shi (时辰)}
\affiliation{Department of Physics, Auburn University, Auburn, AL 36849, USA}
\email{chenshi@auburn.edu} 

\author[0000-0002-6276-7771]{Lorenzo Matteini}
\affiliation{Imperial College London, South Kensington Campus, London SW7 2AZ, UK}
\email{l.matteini@imperial.ac.uk} 

\author[0000-0002-4625-3332]{Trevor A. Bowen}
\affiliation{Space Sciences Laboratory, University of California, Berkeley, CA 94720-7450, USA}
\email{tbowen@berkeley.edu}

\author[0000-0003-4177-3328]{Alfred Mallet}
\affiliation{Space Sciences Laboratory, University of California, Berkeley, CA 94720-7450, USA}
\email{alfred.mallet@berkeley.edu }

\author[0000-0002-7653-9147]{A. Larosa}
\affiliation{Istituto per la Scienza e Tecnologia dei Plasmi (ISTP), Consiglio Nazionale delle Ricerche, I-70126 Bari, Italy}
\email{andrea.larosa@istp.cnr.it}

\author[0000-0003-2880-6084]{Anna Tenerani}
\affiliation{Department of Physics, The University of Texas at Austin, TX 78712, USA}
\email{Anna.Tenerani@austin.utexas.edu} 

\author[0000-0002-7572-4690]{Timothy S. Horbury}
\affiliation{Imperial College London, South Kensington Campus, London SW7 2AZ, UK}
\email{t.horbury@imperial.ac.uk} 

\begin{abstract}
We analyze \emph{Parker Solar Probe} and \emph{Solar Orbiter} measurements to investigate the formation and evolution of magnetic-field reversals (``switchbacks'') across the Alfv\'en surface ($M_a\simeq 1$), where $M_a$ is the Alfv\'en Mach number (ratio of the bulk-flow speed to the Alfv\'en speed). We identify a sub-Alfv\'enic switchback population whose apparent scarcity in earlier studies can be attributed to two diagnostic biases: conditioning on the instantaneous Alfv\'en Mach number, which can be transiently elevated above unity by radial-velocity enhancements that accompany large-amplitude Alfv\'enic rotations, and the use of short-window local-mean backgrounds that partially track these rotations and underestimate the associated deflection angles. When $M_a$ is treated as a bulk-stream property using rolling medians, and deflections are referenced to event-independent backgrounds---either a Parker-spiral direction or a sufficiently long rolling median---sub-Alfv\'enic switchbacks are systematically recovered. Average deflection angles $\langle \theta \rangle$ exhibit two distinct regimes as a function of $M_a$. For $M_a \lesssim 1$, $\langle \theta \rangle$ increases rapidly with $M_a$ and exhibits little dependence on the window used to define the rolling-median background, consistent with expansion-driven amplification of Alfv\'enic fluctuations. For $M_a \gtrsim 1$, the evolution becomes scale dependent: at large scales, $\langle \theta \rangle$ continues to grow with $M_a$ but at a reduced rate relative to the sub-Alfv\'enic regime, whereas at smaller scales the growth saturates, consistent with turbulent decay and dissipation. Collectively, these results indicate that switchbacks need not originate only in the super-Alfv\'enic solar wind. Instead, they are consistent with a formation pathway in which coronal fluctuations are amplified by large-scale expansion through the sub-Alfv\'enic regime, with subsequent propagation into the super-Alfv\'enic wind where turbulent decay modifies their scale-dependent properties.
\end{abstract}



\section{Introduction}\label{sec:intro}
The inner heliosphere is permeated by a wide spectrum of large-amplitude, weakly compressive fluctuations, in which the magnetic-field strength, density, and thermal pressure often exhibit negligible space--time variability ($\delta \rho \approx \delta p \approx \delta|\mathbf{B}| \approx 0$). Velocity--magnetic correlations identify Alfv\'enic \citep{Alfven_1942} disturbances propagating along the magnetic field; the associated Alfv\'enic flux is predominantly directed away from the Sun \citep[e.g.,][]{belcher_large-amplitude_1971, Roberts_1987, Grappin_1990, Tu95}. Driven by photospheric forcing and/or low-coronal reconnection \citep{Tomczyk_2007, McIntosh_2011}, this Alfv\'enic component governs the transport of energy and momentum through the solar wind and therefore constrains both heating and bulk acceleration \citep{Osterbrock_1961, Parker_1965, Verdini_2009ApJ, Chandran_Perez_2019, Perez_2021}.

A conspicuous subset of this spectrum consists of magnetic-field ``switchbacks'' (SBs): impulsive, large-angle rotations of $\mathbf{B}$ relative to a nominal background field, typically taken as the Parker-spiral field or a local running average \citep{2026_SBs_review_badman}. SBs were identified in earlier heliospheric measurements \citep{1966JGR....71.3315M, Balogh_1999, Gosling_2009, Matteini_2014, 2018_horbury_switchbacks}, but \textit{Parker Solar Probe} \citep[PSP;][]{fox_solar_2016} and \textit{Solar Orbiter} \citep[SolO;][]{muller_solar_2020} established them as a defining feature of the near-Sun solar wind \citep{bale_highly_2019, Horbury_2020_SBs, deWit_sbs_WT}.
This motivates renewed scrutiny of the origin and evolution of Alfv\'enic perturbations in the solar wind \citep[e.g.,][]{Tenerani2020, Squire_2020, Mozer2020_SB, Shoda_2021, 2021_larosa, Tenerani_2021, Shi_2022_patches, Squire2022, squire2022b, Johnston2022, Tenerani_23, Larosa_2024, Shi_2024, Choi_2025, Bowen_2025_SBs}: are SBs ``ex-situ'' structures rooted in low-coronal dynamics and advected outward \citep[e.g.,][]{Fisk_2020_SBs, Sterling_2020_SBs, Zank_2020_SBs, Magyar_2021_SBs_solar}, or do they arise ``in-situ'' as the wind expands and becomes turbulent?

In-situ generation mechanisms fall broadly into shear/stream-interaction \citep[e.g.,][]{Landi_2006_SBs, Ruffolo2020, Toth_2023} and expansion-driven amplification of Alfv\'enic fluctuations \citep[e.g.,][]{Squire_2020, Mallet_2021, Shoda_2021, Johnston2022, squire2022b, Squire_2022}.  In the latter picture, SBs arise as a nonlinear outcome of the advection of finite-amplitude Alfv\'enic wavepackets through an inhomogeneous, accelerating background. The fluctuations exchange energy and momentum with the mean flow, so wave energy is not conserved; in the absence of dissipation, the radial evolution is instead constrained by wave-action conservation. In steady state, the wave-action density $\mathcal{S}=\mathcal{E}/\omega_0$ (with $\omega_0$ the intrinsic frequency in the plasma frame) is an adiabatic invariant along ray trajectories \citep{witham1965general, bretherton1968wavetrains, Heinemann_Olbert, Velli_93, 2015_Chandran_wave_action, Tenerani_EBM}.

Writing the fluctuation energy density as $\mathcal{E}=(\rho/4)\big[|\mathbf{z}^{+}|^{2}+|\mathbf{z}^{-}|^{2}\big]$ in terms of Elsasser variables $\mathbf{z}^{\pm}=\mathbf{v}\pm \mathbf{b}/\sqrt{\mu_0\rho}$ \citep{elsasser_1950}, and considering outward-dominated propagation ($|z^{+}|\gg|z^{-}|$), wave-action conservation combined with conservation of mass and magnetic flux implies a characteristic $M_a$-dependence for the normalized fluctuation amplitude, derived in Appendix~\ref{app:wkb_theta_ps} (Section~\ref{app:wkb_theta_ps:wkb}),
\begin{equation}
\left(\frac{\delta B_{\mathrm{rms}}}{B_0}\right)^2 \;\propto\; \frac{M_a^{3}}{(M_a+1)^2}.
\end{equation}
where $M_a\equiv U/V_A$, and the bulk-stream and instantaneous definitions used in this work are given in Appendix~\ref{Appendix:Methods} (Section~\ref{app:Ma_defs}). Because $M_a$ typically increases with heliocentric distance as the wind accelerates and $V_A$ decreases, this scaling implies monotonic, formally unbounded amplification of $(\delta B_{\mathrm{rms}}/B_0)^2$ with $M_a$ in the undamped WKB limit \citep{Hollweg_1974, Mallet_2021}. As $\delta B/B_0$ approaches unity, $\delta|\mathbf{B}|\approx 0$ implies primarily directional fluctuations: $\mathbf{B}$ undergoes large deflections at nearly fixed magnitude
(spherical polarization; \citealt{Goldstein_1974, Barnes_1981, squire2022b, Matteini_2024}). Relative to a fixed $B_0$, such deflections generically require a finite $\delta B_{\parallel}$ to maintain $\delta|\mathbf{B}|\approx 0$ (purely transverse large-amplitude solutions are special cases, e.g., circular polarization). For common geometries, $\delta B_{\parallel}\sim\mathcal{O}(B_0)$ permits strong rotations and can produce local polarity reversals \citep{barnes_large-amplitude_1974, Mallet_2021}. In this sense, SBs can be the nonlinear continuation of initially modest Alfv\'enic perturbations evolving under wave-action constraints in an expanding background \citep{Squire_2020, Mallet_2021, Squire2022, squire2022b, Johnston2022}.

Expansion-driven amplification operates in tandem with a turbulent cascade \citep{iroshnikov_turbulence_1963, kraichnan_inertial-range_1965, goldreich_toward_1995, chandran_intermittency_2015, Mallet_2017, Chandran_2025} that can both facilitate and limit the formation of SBs \citep{Squire2022, Johnston2022}. A developed turbulent cascade requires counter-propagating components ($z^{-}$), generated by reflection on large-scale gradients, parametric decay, or velocity shear \citep{Heinemann_Olbert, velli_turbulent_1989, Galeev_1963, Roberts_1987, Goldstein_1989, 1997_prunareti_PDI}. The efficiency of nonlinear transfer is governed by the nonlinearity parameter $\chi\equiv\tau_A/\tau_{nl}\sim (k_\perp \delta B)/(k_\parallel V_A)$. Recent modeling indicates a regime dependence: in the super-Alfv\'enic wind, increasing spectral obliquity drives $\chi\gtrsim 1$, producing strong, scale-dependent decay that can arrest further expansion-driven growth \citep{Chandran_Perez_2019, Johnston2022, Tenerani_2021}; in the sub-Alfv\'enic corona, rapid acceleration maintains $\chi\lesssim 1$, leaving a window in which expansion forcing outpaces turbulent damping \citep{Johnston2022, Squire2022}. In-situ observations support this framework: while sub-Alfv\'enic fluctuations exhibit near-adiabatic evolution with weak dissipation \citep{2025_Sioulas_AWs}, the super-Alfv\'enic wind displays the strong, scale-dependent decay characteristic of a developed turbulent cascade \citep{Villante_1982, bavassano_radial_1982, Roberts_1987, Horbury_2001, Tenerani_2021}.

\begin{figure}
\centering
\includegraphics[width=0.47\textwidth]{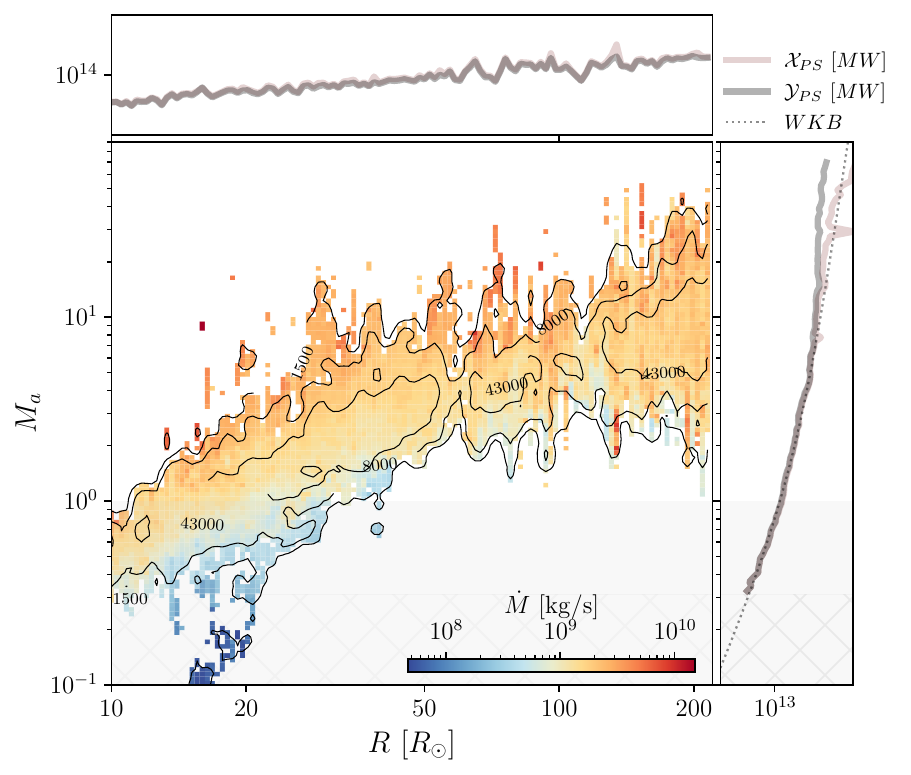}
\caption{Mass-flux proxy $\dot{M}$ in the $(R,M_a)$ plane, where $R\equiv r/R_{\odot}$ and $r$ is heliocentric distance. Colors show the binned mean $\langle \dot{M}\rangle$ (log scale), and black contours indicate the sample-count levels used to delineate regions with sufficient statistics. The top and right margins show one-dimensional profiles of the $M_a$-conditioned diagnostics (definitions in Appendix~\ref{app:wkb_theta_ps:diagnostics}): $\mathcal{Y}_{PS}\equiv \dot{M}\,2\big(U^{\mathrm{Bulk}}\big)^2\!\left(1-\langle \cos\Theta_{PS}\rangle_{w}\right)$ (black) and $\mathcal{X}_{PS}\equiv \dot{M}\,\big(U^{\mathrm{Bulk}}\big)^2(\delta B_{PS}/B_0)^2$ (colored), averaged over $M_a$ (top) and over $R$ (right), respectively. The dotted curve shows the best-fit WKB  profile $\alpha\,g(M_a)$ with $g(M_a)=M_a^3/(M_a+1)^2$ over the fitted $M_a$ range.}
\label{fig:massflux_heatmap}
\end{figure}

Despite this theoretical and observational consistency, a critical tension remains. If the sub-Alfv\'enic regime supports amplification toward $\delta B/B_0\sim 1$, then large-angle rotations can occur below the Alfv\'en surface. Several in-situ studies instead reported a disappearance of SBs in sub-Alfv\'enic intervals \citep{Bandyopadhyay_alfven, Akhavan_Tafti_2024_BS, Jagarlamudi_2023}, an apparent ``dropout'' that has been interpreted as evidence for SB formation confined to the super-Alfv\'enic wind, for example via shear-driven processes acting at or beyond the Alfv\'en surface \citep[e.g.,][]{Ruffolo2020, Schwadron_McComas_2021_SBs}.\footnote{Remote-sensing observations reported a propagating S-shaped structure interpreted as evidence for SB-like geometry in the corona \citep{Telloni_2022}.}

Here we show that the reported deficit of sub-Alfv\'enic switchbacks is a diagnostic artifact tied to both the definition of $M_a$ and the choice of background field (Appendix~\ref{Appendix:Methods}). Large-angle Alfv\'enic rotations are coupled to transient radial-velocity enhancements \citep{Matteini_2014}, which can raise the \emph{instantaneous} local $M_a$ above unity even when the underlying stream remains sub-Alfv\'enic. Conditioning deflection statistics on instantaneous $M_a$ therefore transfers intrinsically sub-Alfv\'enic, SB-like intervals into super-Alfv\'enic bins. In addition, when the ``background'' field is defined by a short running average, it partially follows the rotation itself, reducing the measured deflection and biasing SB occurrence low by construction. When $M_a$ is instead treated as a bulk-stream property (via a rolling-median estimate) and deflections are measured relative to an event-independent background (e.g., a sufficiently long rolling median or a Parker-spiral reference), a non-negligible sub-Alfv\'enic SB population is recovered.

With these diagnostic choices fixed, the inferred trends follow the expectations for expansion-driven amplification. Inside the Alfv\'en surface, the mean deflection angle increases rapidly with $M_a$ and exhibits little scale dependence, consistent with wave-action conservation in a weakly dissipative regime. Beyond the Alfv\'en surface, the growth becomes scale dependent: turbulent dissipation limits small-scale deflections, while larger-scale rotations continue to grow.

The paper is organized as follows. Section~\ref{sec:Statistical_Results} presents the statistical dependence of deflection angles on $M_a$ and scale. Section~\ref{sec:Discussion} interprets these findings in the context of wave-action conservation and turbulence. Section~\ref{sec:Conclusions} summarizes the implications for switchback generation models. Data selection and methods are detailed in the Appendices.

\begin{figure*}
\centering
\setlength\fboxsep{0pt}
\setlength\fboxrule{0.0pt}
\includegraphics[width=1\textwidth]{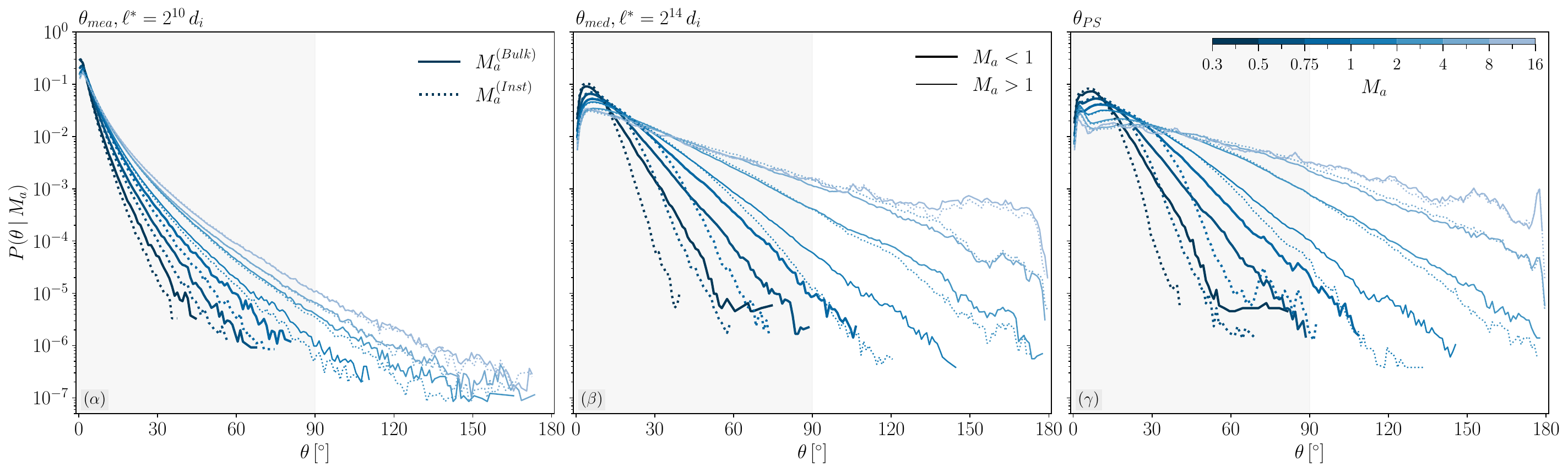}
\caption{Conditional PDFs $P(\theta\,|\,M_a)$ of the magnetic-field deflection angle $\theta$ (definitions in Appendix~\ref{Appendix:Methods}, Sections~\ref{app:Ma_defs} and \ref{app:def_angles}). Panels~$(\alpha)$ and $(\beta)$ measure $\theta$ relative to a local background field defined by a dynamic window of scale $\ell^\ast$: Panel~$(\alpha)$ uses a local-mean background ($\theta_{\mathrm{mea}}$, $\ell^\ast=2^{10}\,d_i$) and Panel~$(\beta)$ uses a local-median background ($\theta_{\mathrm{med}}$, $\ell^\ast=2^{14}\,d_i$). Panel~$(\gamma)$ measures $\theta$ relative to the Parker-spiral direction ($\theta_{PS}$) as the reference. Color encodes the $M_a$ bin (top color bar). Solid (dotted) curves condition on the bulk-stream (instantaneous) Mach number, $M_a^{\mathrm{Bulk}}$ ($M_a^{\mathrm{Inst}}$). Curve thickness denotes $M_a<1$ (thin) and $M_a>1$ (thick) using the same Mach-number definition used to condition each curve. Shaded regions mark $\theta<90^\circ$.}\label{fig:def_angles_pdf}
\end{figure*}

\section{Results}\label{sec:Statistical_Results}

\subsection{Parker-spiral--referenced normalized amplitudes and deflection angles}\label{subsec:PS_WKB_test}

Figure~\ref{fig:massflux_heatmap} compares the deflection-based diagnostic $\mathcal{Y}_{PS}$ and the normalized-amplitude diagnostic $\mathcal{X}_{PS}$, defined in Appendix~\ref{app:wkb_theta_ps:diagnostics}, using a Parker-spiral baseline. The main panel reports the sample coverage in the $(R,M_a)$ plane, colored by the binned mean mass-flux proxy $\langle\dot{M}\rangle$ (log scale), with black contours marking the sample-count levels used to delineate regions with sufficient statistics. Hatched regions denote the lowest $0.1\%$ of the $\dot{M}$ distribution; this mask suppresses the influence of the extreme low-$\dot{M}$ tail on low-count bins, which is relevant because both $\mathcal{X}_{PS}$ and $\mathcal{Y}_{PS}$ carry an explicit linear scaling with $\dot{M}$ \citep{2025_Sioulas_AWs}.

We reference both $\Theta_{PS}$ and $\delta B_{PS}$ to the Parker-spiral field $\mathbf{B}_{PS}$, specified independently of the windowed fluctuations used to define the diagnostics. Near the Sun, polarity-sector fluctuations can be strongly one-sided, with the sector-anchoring component fluctuating about a base value rather than symmetrically about a windowed mean \citep{Gosling_2009}. In that regime, a rolling-mean background constructed from the same window is not an external reference: it is biased toward the dominant excursion and can partially co-rotate with it, reducing both the inferred deflection and the inferred normalized amplitude. Referencing to $\mathbf{B}_{PS}$ removes this self-referential attenuation by keeping the baseline decoupled from the rotations being quantified.

The marginal profiles summarize the behavior of the Parker-spiral--referenced diagnostics under conditioning by $M_a$ and averaging over $R$. The $R$-averaged curves $\langle\mathcal{X}_{PS}\rangle_R$ and $\langle\mathcal{Y}_{PS}\rangle_R$ lie close to one another over most of the sampled $M_a$ range, indicating that the deflection-based and normalized-amplitude measures track each other under the same reference. In the limit of spherically polarized Alfv\'enic perturbations with $\delta|{\bf B}|\ll|\delta{\bf B}|$, Eq.~\eqref{eq:angle_amplitude_mapping} provides a direct mapping between $\Theta_{PS}$ and $\delta B_{PS}/B_0$; the observed agreement in the marginal means is consistent with that geometric linkage operating over the bulk of the sample \citep[see also][]{Larosa_2024}. The $M_a$-conditioned means follow the best-fit WKB profile $\alpha\,g(M_a)$ over the fitted range (Appendix~\ref{app:wkb_theta_ps}), with systematic departures confined to the largest $M_a$. Such high-$M_a$ departures occur in the portion of parameter space where compressive contributions can become more prominent in near-Sun intervals \citep{Tenerani_2021, Larosa_2024}, and therefore motivate an explicit compressibility check when interpreting deviations from the WKB trend.

\subsection{PDFs of Deflection Angles}\label{subsec:deflection_angle_PDFs}

Figure~\ref{fig:def_angles_pdf} shows the conditional PDFs $P(\theta\,|\,M_a)$ of magnetic-field deflections. The angle definitions, reference fields, and the bulk/instantaneous Mach numbers are given in Appendix~\ref{Appendix:Methods} (Sections~\ref{app:Ma_defs} and \ref{app:def_angles}). Panels~$(\alpha)$--$(\beta)$ measure $\theta$ relative to local backgrounds defined by a dynamic smoothing window of scale $\ell^\ast$, whereas Panel~$(\gamma)$ uses the Parker-spiral direction ($\theta_{PS}$) as the reference. Panel~$(\alpha)$ adopts a local mean ($\theta_{\mathrm{mea}}$; $\ell^\ast=2^{10}\,d_i$) and Panel~$(\beta)$ adopts a local median ($\theta_{\mathrm{med}}$; $\ell^\ast=2^{14}\,d_i$). These panels therefore compare representative choices of both estimator (mean versus median) and window scale; the difference should not be interpreted as an estimator-only effect.

Two analysis choices control the high-$\theta$ tail ($\theta\gtrsim 90^\circ$). To isolate the conditioning effect, compare the solid and dotted curves at fixed color (i.e., at fixed $M_a$ bin). The separation is most apparent in Panels~$(\beta)$ and $(\gamma)$ near the $M_a\simeq 1$ transition (thick curve(s), indicating $M_a\leq 1$), where the dotted curves show a pronounced deficit at $\theta\gtrsim 90^\circ$. Conditioning on the instantaneous Mach number ($M_a^{\mathrm{Inst}}$; dotted) suppresses the sub-Alfv\'enic tail because Alfv\'enic rotations are coupled to radial-velocity enhancements \citep{Matteini_2014}. Those transient increases in $V_r$ can drive $M_a^{\mathrm{Inst}}$ across unity during the rotation itself, reassigning near-critical, switchback-like intervals from sub- to super-Alfv\'enic bins. Conditioning instead on the bulk-stream Mach number ($M_a^{\mathrm{Bulk}}$; solid) is insensitive to these short excursions and retains a population of reversals ($\theta\gtrsim 90^\circ$) below the Alfv\'en critical surface.

The inferred tail also depends on how the background is estimated when intermittent rotations occur within a finite window. A local mean is pulled toward large-amplitude outliers inside the window, rotating the estimated background direction toward the event and reducing the inferred deflection angle. This attenuation is strongest when $\ell^\ast$ is comparable to the event duration. A rolling median is robust to such outliers and weakens the coupling between intermittent excursions and the estimated background direction, which preserves the high-$\theta$ tail relative to local-mean choices at smaller $\ell^\ast$.

These methodological effects can produce an apparent deficit of sub-Alfv\'enic, switchback-like reversals reported in recent work \citep[e.g.,][]{Bandyopadhyay_alfven, Akhavan_Tafti_2024_BS, adhikari2025_Transalfv}. With bulk conditioning and robust background estimation, we identify $\mathcal{O}(10^2)$ SBs within sub-Alfv\'enic streams, using the identification method outlined in Appendix~\ref{Appendix:Methods}, Section~\ref{app:def_angles}.

Across all panels, the PDFs broaden systematically with increasing $M_a$, consistent with a monotonic increase in the occurrence of large-angle deflections as the flow crosses from sub- to super-Alfv\'enic. The same statistics binned by heliocentric distance $r$ (not shown) exhibit considerably weaker ordering, with only modest tail broadening that becomes apparent mainly for the smoothest background definitions. This comparison indicates that $M_a$ organizes the deflection statistics more effectively than $r$ alone.

\subsection{Average Deflection Angles}\label{subsec:deflection_angles}

\begin{figure}[t]
\centering
\includegraphics[width=0.4\textwidth]{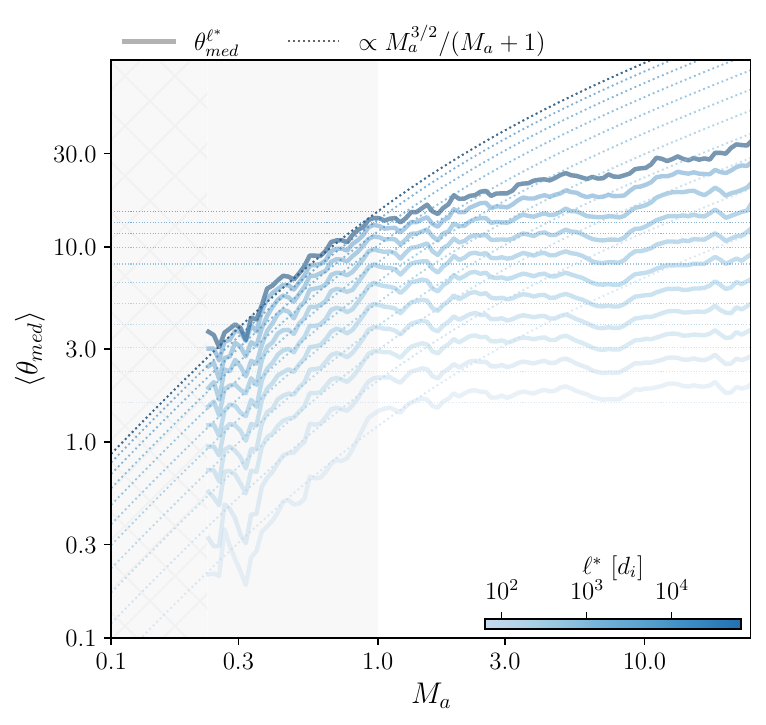}
\caption{Average deflection angle $\langle \theta_{\mathrm{med}} \rangle$ as a function of  $M_a$. Colored curves indicate deflection angles estimated relative to a local, scale-dependent background magnetic field $\mathbf{B}^{\mathrm{med}}_{\ell^\ast}(t)$, with the scale $\ell^\ast$ parametrized by the smoothing window width shown in the colorbar (definitions in Appendix~\ref{Appendix:Methods}, Section~\ref{app:def_angles}).}
\label{fig:Average_def_angles}
\end{figure}

Figure~\ref{fig:Average_def_angles} plots $\langle\theta_{\mathrm{med}}^{\ell^\ast}\rangle$ versus $M_a$, where $\theta_{\mathrm{med}}^{\ell^\ast}$ is measured relative to a rolling-median background evaluated on window scale $\ell^\ast$ (Appendix~\ref{Appendix:Methods}, Section~\ref{app:def_angles}). The dotted curves indicate the reference WKB scaling $\propto M_a^{3/2}/(M_a+1)$.

The curves change character across $M_a\simeq 1$. For $M_a\lesssim 1$, the $M_a$-dependence collapses across $\ell^\ast$ (weak scale dependence), with $\ell^\ast$ mainly setting the normalization. This behavior is consistent with a regime in which dissipation is weak across the set of resolved windows, so the deflection statistics on all $\ell^\ast$ evolve at comparable rates. In the super-Alfv\'enic regime ($M_a\gtrsim 1$), the behavior becomes strongly scale dependent. Small-window curves flatten at $\langle \theta_{\mathrm{med}}\rangle=\mathcal{O}(1)$, whereas large-window curves continue to increase with $M_a$ but with reduced slope. At fixed $M_a$, $\langle \theta_{\mathrm{med}}\rangle$ increases monotonically with $\ell^\ast$, and the separation between window choices widens across the super-Alfv\'enic range.

For $M_a\gtrsim 1$, the window dependence strengthens: small-$\ell^\ast$ curves roll over and approach a plateau, whereas large-$\ell^\ast$ curves continue to increase with $M_a$ with reduced slope, and the separation between windows grows with $M_a$. The observed scale dependence for $M_a>1$ is consistent with the broader scale-dependent evolution of turbulent fluctuations in the expanding wind \citep{Tenerani_2021, Tenerani_23}: smaller-scale fluctuations decay more rapidly than outer-scale rotations. It is also consistent with reports that, at small scales, the statistical properties exhibit only weak radial ordering \citep{Sioulas_2022_intermittency, Larosa_2024, Sioulas_2023_anisotropic}, plausibly because the shorter nonlinear times $\tau_{nl}$ at those scales drive a more rapid approach to a quasi-stationary state than at larger scales.

\section{Discussion}\label{sec:Discussion}

\subsection{Influence of Conditioning on Switchback Statistics}\label{subsec:conditioning_bias}

A central result of this study is the identification of a systematic bias introduced when conditioning switchback statistics on the instantaneous Alfv\'en Mach number, $M_a^{\mathrm{Inst}}(t)$ (Appendix~\ref{Appendix:Methods}, Section~\ref{app:Ma_defs}). This bias arises from the intrinsic kinematic coupling between magnetic deflections and radial velocity in large-amplitude, approximately constant-magnitude Alfv\'enic fluctuations.

As detailed by \citet{Matteini_2014}, spherical polarization of the magnetic field vector ($|\mathbf{B}| \approx \text{const}$) implies that deviations from the background field direction are coupled to variations in the proton radial velocity, $v_R$. The velocity variation associated with such deflections is captured by the expression \citep{Matteini_2014}
\begin{equation}
\frac{\delta v_R}{v_{\text{wave}}} = \pm \frac{\delta B_R}{B} = \pm \left( \cos(\theta_{BR}) - \cos(\theta_0) \right),
\end{equation}
where $\theta_{BR}$ and $\theta_0$ follow the notation of \citet{Matteini_2014}, $v_{\text{wave}} \sim v_A \sqrt{R_A}$, and $R_A$ is the ratio of kinetic to magnetic fluctuation energy. For states approaching pure Alfv\'enicity ($R_A \to 1$), large-angle deflections (where $\cos \theta < \cos \theta_0$) produce positive excursions in the radial velocity, $\delta v_R > 0$.

This coupling has critical implications for statistical binning near the Alfv\'en critical point. In a nominally sub-Alfv\'enic stream ($M_{a, \text{bulk}} \lesssim 1$), a large-amplitude switchback-like rotation generates a localized velocity enhancement that may transiently drive the instantaneous Mach number above unity ($M_{a, \text{inst}} > 1$). Consequently, conditioning analysis on $M_{a, \text{inst}}$ can reclassify these large-deflection events into super-Alfv\'enic bins.

This distinction points to a deeper issue: while an instantaneous $M_a^{\mathrm{Inst}}(t)$ is well-defined, it is not a robust regime label on switchback scales for conditioning, because it is strongly modulated by the same Alfv\'enic fluctuations whose statistics are being measured. The Alfv\'en Mach number $M_a = U / V_A$ is a bulk property of the plasma stream and reflects the large-scale competition between advection and magnetic tension. Although transient deflections may modulate the local speed over short intervals, they do not imply that adjacent plasma parcels experience fundamentally different Mach-number regimes. The solar wind evolves as a continuous medium; arbitrarily close regions do not independently transition between sub- and super-Alfv\'enic states. Conditioning statistical analysis on the instantaneous value of $M_a$---especially on timescales comparable to fluctuation durations---therefore risks misclassification, reallocating structures based on instantaneous excursions rather than underlying stream properties.

\section{Conclusions}\label{sec:Conclusions}

Using PSP and SolO measurements, we re-examine the observational basis of the reported ``sub-Alfv\'enic switchback dropout'' \citep{Bandyopadhyay_alfven, Akhavan_Tafti_2024_BS, Jagarlamudi_2023}. In this dataset, an apparent deficit follows from two compounded methodological effects. First, conditioning on an \emph{instantaneous} $M_a$ can relabel intervals as super-Alfv\'enic when large-amplitude Alfv\'enic rotations coincide with radial-velocity enhancements. Second, reference fields defined by short running averages partially track intermittent deflections, suppressing the measured angles. When $M_a$ is treated as a bulk-stream property and deflections are referenced to event-independent backgrounds (long rolling medians and/or a Parker-spiral direction), large-angle rotations and polarity reversals are recovered systematically for $M_a<1$. 

A practical element of this work is the use of a Parker-spiral direction as an event-independent background for measuring deflections and constructing fluctuation diagnostics. This choice is motivated by the one-sided, outward-dominated Alfv\'enic component of the near-Sun wind and by the observed association between large rotations and radial-velocity enhancements \citep{Gosling_2009, Matteini_2014}. In the RD-like, nearly constant-$|B|$ limit adopted in Appendix~\ref{app:fluctuation_decomposition},  (Eq.~\eqref{eq:BPS_amp_def}), the Parker-spiral geometry provides the mapping in Eq.~\eqref{eq:angle_amplitude_mapping} between $1-\langle\cos\Theta_{PS}\rangle_w$ and $(\delta B_{PS}/B_0)^2$ on the analysis window. This permits an internal cross-check of deflection-based and amplitude-based measures and a direct comparison to WKB expectations for the $M_a$-profile.

After controlling conditioning and background biases, the deflection statistics organize most cleanly by $M_a$, with a regime separation near $M_a\simeq 1$. For $M_a\lesssim 1$, the mean deflection increases rapidly with weak scale dependence across the analyzed windows, matching WKB-like, expansion-driven amplification of an outward-dominated Alfv\'enic component in a weakly dissipative regime. The Parker-spiral--referenced conditioned profiles recover the expected wave-action profile $g(M_a)$ under the rotation-dominated approximation stated in Appendix~\ref{app:wkb_theta_ps}, placing the sub-Alfv\'enic growth on the same footing as WKB evolution in the outer corona and near-Sun wind \citep{Heinemann_Olbert, Velli_93, 2015_Chandran_wave_action}. For $M_a\gtrsim 1$, the evolution becomes scale dependent: smaller-scale deflections saturate while larger-scale deflections continue to increase more gradually, matching a super-Alfv\'enic regime in which turbulent cascade preferentially limits smaller-scale fluctuations \citep{Tenerani_2021, Larosa_2024, Sioulas_2022_intermittency, Sioulas_2023_anisotropic}.

These results constrain switchback-generation scenarios. Mechanisms that confine switchback formation to $M_a\gtrsim 1$ \citep[e.g.,][]{Ruffolo2020, Schwadron_McComas_2021_SBs} are disfavored in the specific sense that large-angle rotations and polarity reversals are present in sub-Alfv\'enic streams once regime labeling and reference-field definitions are made independent of the measured fluctuations. By contrast, models in which finite-amplitude coronal Alfv\'enic fluctuations are amplified by expansion are favored by the observed sub-Alfv\'enic growth and by the Parker-spiral--referenced $g(M_a)$ dependence \citep{Squire_2020, Mallet_2021, Johnston2022, Squire2022, squire2022b}. This does not exclude additional contributions from other processes; it establishes that a formation pathway restricted to the super-Alfv\'enic regime is not required to account for the emergence and $M_a$-ordered evolution of large-angle deflections in the present sample.

Overall, the observations are compatible with a picture in which coronal Alfv\'enic fluctuations undergo expansion-driven amplification through the sub-Alfv\'enic corona, producing switchback-like rotations, and are subsequently reshaped in the super-Alfv\'enic wind by scale-dependent turbulent decay.

\begin{acknowledgments}

This research was funded in part by the FIELDS experiment
on the Parker Solar Probe spacecraft, designed and developed under NASA contract
NNN06AA01C; the NASA Parker Solar Probe Observatory Scientist
grant NNX15AF34G and the  HERMES DRIVE NASA Science Center grant No. 80NSSC20K0604.
The instruments of PSP were designed and developed under NASA contract NNN06AA01C. NS was supported by NASA Award 80NSSC24K0272.
CS was supported by NASA ECIP 80NSSC23K1064.

\software{The authors acknowledge the following open-source packages: \citetalias{van1995python}, \citetalias{2020SciPy-NMeth}, \citetalias{mckinney2010data}, \citetalias{Hunter2007Matplotlib}, \citetalias{angelopoulos_space_2019}, \citetalias{MHDTurbPy_Sioulas}.}
\end{acknowledgments}

\clearpage
\appendix

\section{Data Selection \& Processing}\label{App:Data_Sel_and_Processing}

\subsection{Parker Solar Probe (\citetalias{fox_solar_2016})}\label{Appendix_Data:PSP}

We analyze \citetalias{fox_solar_2016} observations spanning 1~October~2018 to 23~June~2025, covering the first 24 perihelion encounters (E1--E24).
Magnetic-field measurements are taken from Level~2 Fluxgate Magnetometer (FGM) data products \citep{bale_fields_2016} in RTN coordinates.
Proton plasma moments are obtained from Level~3 Solar Wind Electrons Alphas and Protons (SWEAP) data.
For heliocentric distances $R \gtrsim 0.25$~au, moments are taken from the Solar Probe Cup (SPC; \citealt{kasper_solar_2016}); for smaller $R$, we use the Solar Probe Analyzer (SPAN), also part of SWEAP.

Following SWEAP instrument-team recommendations, we form 10-minute rolling medians of the SPAN-i proton density and remove intervals with abrupt density drops exceeding $50\%$ relative to the local rolling-median baseline.
To mitigate partial obstruction by the spacecraft heat shield, we further require that the azimuthal flow angle in the instrument frame remain below approximately $165^\circ$.

Rather than adopting proton number densities $n_p$ directly from plasma moments, we use electron densities from quasi-thermal noise (QTN) spectroscopy performed by the FIELDS instrument \citep{Moncuquet_2020}.
Assuming charge neutrality and an alpha-particle abundance $a_\alpha\equiv n_\alpha/n_p=0.04$, so that $n_e=n_p+2n_\alpha=n_p(1+2a_\alpha)=1.08\,n_p$, we convert $n_e$ to $n_p$ via $n_p=n_e/1.08$.
We exclude intervals in which the SPAN-i proton density differs from the corresponding QTN estimate by more than an order of magnitude; this is used as a coarse consistency screen to remove pathological plasma-moment intervals rather than to cross-calibrate the two estimates.
All measurements at heliocentric distances $R \geq 100\,R_{\odot}$ are excluded due to degraded performance/coverage of the adopted plasma and QTN products in that regime.

\subsection{Solar Orbiter (\citetalias{muller_solar_2020})}\label{Appendix_Data:SolO}

We analyze \citetalias{muller_solar_2020} measurements spanning 1~July~2020 to 1~March~2023.
Magnetic-field data are provided by the Magnetometer (MAG) instrument \citep{horbury_solar_2020}, and ion bulk properties are obtained from the Proton and Alpha Particle Sensor (SWA-PAS) within the Solar Wind Analyser (SWA) suite \citep{owen_solar_2020}.
Consistent with the procedure adopted for \citetalias{fox_solar_2016}, the proton number density $n_p$ is inferred from electron number densities determined via QTN measurements.

\subsection{Data Processing}\label{Appendix_methods:Data_processing}

The dataset is partitioned into contiguous segments of duration $D=24\,\mathrm{h}$, with successive segments overlapping by $4\,\mathrm{h}$.
Only segments containing QTN measurements are retained.
A segment is discarded if the fraction of missing samples exceeds $2\%$ for the magnetic-field series, $10\%$ for the velocity series, or $50\%$ for the QTN series.
The segment duration exceeds the longest time-domain windows used in the analysis, and the overlap limits edge losses at segment boundaries.

Plasma-moment time series are preprocessed with a Hampel filter \citep{davies_identification_1993}, applied over a 300-sample sliding window, to remove outliers exceeding three local standard deviations.
Segments are then screened for heliospheric current sheet (HCS) crossings, coronal mass ejections (CMEs), and other prominent transients by manual inspection of time series and summary plots of the magnetic field and plasma moments.
For each identified event, a temporal buffer of $10\,\mathrm{min}$ is imposed on either side, and the buffered region is excised.
The time stamps of all excised intervals are retained as part of the analysis metadata.

After transient excision, each time series is re-examined to identify individual data gaps.
Sub-segments are excluded if any single gap exceeds $30\,\mathrm{s}$ in the magnetic-field measurements, $1\,\mathrm{min}$ in the velocity measurements, or $5\,\mathrm{min}$ in the QTN-derived proton density.
Remaining gaps are filled by linear interpolation only when permitted by these per-gap thresholds.
After computation of the target quantities, additional buffers equal to $20\%$ of the respective gap thresholds are applied before and after each identified gap, and the corresponding samples are removed to suppress edge-related artefacts in subsequent analyses.

\section{Methods and Definitions}\label{Appendix:Methods}

\subsection{Conventions and window operators}\label{app:operators_units}

We use two time-domain aggregation operators and one bin-average operator.
For a scalar or vector quantity $q(t)$, the centered rolling mean over a window of width $w$ is
\begin{equation}
\langle q\rangle_{w}(t)\equiv \frac{1}{N_w(t)}\sum_{t'\in \mathcal{W}(t)} q(t'),
\label{eq:window_average_def_app}
\end{equation}
where $\mathcal{W}(t)$ is the set of samples within the window centered at $t$ and $N_w(t)$ is the number of samples in that window.
The corresponding rolling median is denoted $\mathrm{med}_{w}[q](t)$; for vector-valued $q$, the operators are applied component-wise.

Conditioned profiles (e.g., binned in $M_a$) are computed as arithmetic means over a sample set (bin) $\mathcal{S}$,
\begin{equation}
\langle Q\rangle_{\mathcal{S}} \equiv \frac{1}{N_{\mathcal{S}}}\sum_{t\in\mathcal{S}} Q(t),
\label{eq:bin_average_def_app}
\end{equation}
with $N_{\mathcal{S}}$ the number of samples in $\mathcal{S}$.
When conditioning specifically on $M_a$, we denote the resulting bin-average by $\langle\cdot\rangle_{M_a}$ (see \S~\ref{app:wkb_theta_ps:diagnostics}).

\subsection{Bulk and instantaneous Alfv\'en Mach numbers}\label{app:Ma_defs}

We define a large-scale background magnetic field using a fixed window $w_M$,
\begin{equation}
\mathbf{B}_0(t)\equiv \mathrm{med}_{w_M}\!\left[\mathbf{B}\right]\!(t),
\qquad
B_0(t)\equiv |\mathbf{B}_0(t)|,
\qquad
\hat{\mathbf{b}}_0(t)\equiv \mathbf{B}_0(t)/B_0(t).
\label{eq:B0_def_app}
\end{equation}

The analysis was repeated using rolling means for the large-scale magnetic field, yielding similar trends (not shown). Bulk-stream quantities use the same $w_M$; in particular,
$$
n_p^{\mathrm{Bulk}}(t)\equiv \mathrm{med}_{w_M}[n_p](t),\qquad
\mathbf{V}_p^{\mathrm{Bulk}}(t)\equiv \mathrm{med}_{w_M}[\mathbf{V}_p](t),\qquad
U^{\mathrm{Bulk}}(t)\equiv \mathrm{med}_{w_M}[U](t),
$$
where $U(t)\equiv|\mathbf{V}_p(t)|$ and $\mathbf{V}_p(t)$ is the measured proton bulk velocity.

The alpha abundance parameter is $a_\alpha\equiv n_\alpha/n_p$.
The corresponding bulk and instantaneous mass densities are
\begin{equation}
\rho^{\mathrm{Bulk}}(t)\equiv m_p\,n_p^{\mathrm{Bulk}}(t)\,(1+4a_\alpha),
\qquad
\rho(t)\equiv m_p\,n_p(t)\,(1+4a_\alpha).
\label{eq:rho_bulk_inst_def_app}
\end{equation}

The bulk-stream Alfv\'en Mach number is
\begin{equation}
M_a^{\mathrm{Bulk}}(t)\equiv \frac{U^{\mathrm{Bulk}}(t)}{V_A^{\mathrm{Bulk}}(t)},
\label{eq:Ma_bulk_def_app}
\end{equation}
where $V_A^{\mathrm{Bulk}}(t) \equiv B_0(t)\big(\mu_0\,\rho^{\mathrm{Bulk}}(t)\big)^{-1/2}$ is the background Alfv\'en speed under the density convention in Eq.~\eqref{eq:rho_bulk_inst_def_app}.
For diagnosing pointwise conditioning biases, we also define an instantaneous normalization,
\begin{equation}
M_a^{\mathrm{Inst}}(t)\equiv \frac{U(t)}{V_A^{\mathrm{Inst}}(t)},
\label{eq:Ma_inst_def_app}
\end{equation}
where $V_A^{\mathrm{Inst}}(t) \equiv |\mathbf{B}(t)|\big(\mu_0\,\rho(t)\big)^{-1/2}$.
Unless stated otherwise, all $M_a$-conditioned statistics in the main text use $M_a^{\mathrm{Bulk}}$; $M_a^{\mathrm{Inst}}$ is used only to quantify the bias introduced by pointwise conditioning (Section~\ref{subsec:deflection_angle_PDFs}).

\subsection{Smoothing windows normalized to an inertial-length proxy}\label{app:kinetic_windows}

Smoothing scales are specified in units of an inertial-length constructed from the same background conventions used for $M_a$,
\begin{equation}
d_i(t)\equiv \frac{V_A^{\mathrm{Bulk}}(t)}{\Omega_i(t)},
\label{eq:di_Omega_def_app_revised}
\end{equation}
where $\Omega_i(t)=e\,B_0(t)/m_p$ and $B_0(t)\equiv |\mathbf{B}_0(t)|$.

Given a prescribed scale $\ell^\ast \equiv N\,d_i$, the corresponding time-domain window width is set by a relative sweep speed,
\begin{equation}
w_{\ell^\ast}(t) \equiv \frac{N\,d_i(t)}{U_{\mathrm{eff}}(t)}.
\label{eq:w_of_lstar_def_revised}
\end{equation}

We define $U_{\mathrm{eff}}(t)\equiv |\mathbf{U}_{\mathrm{eff}}(t)|$ with
\begin{equation}
\mathbf{U}_{\mathrm{eff}}(t)
=
\mathbf{V}_p^{\mathrm{Bulk}}(t)
-
\mathbf{V}_{sc}(t)
+
\sigma(t)\,V_A^{\mathrm{Bulk}}(t)\,\hat{\mathbf{b}}_0(t),
\label{eq:Ueff_def_app_revised}.
\end{equation}
 The polarity factor $\sigma(t)$ selects the sign corresponding to anti-sunward propagation along the local mean field. In RTN coordinates this may be taken as $\sigma(t)=\mathrm{sgn}(B_0^{r}(t))$, whereas in general geometry it is $\sigma(t)=\mathrm{sgn}\!\big(\hat{\mathbf{b}}_0(t)\!\cdot\!\hat{\mathbf{r}}(t)\big)$. In intervals with $|\hat{\mathbf{b}}_0\!\cdot\!\hat{\mathbf{r}}|\ll 1$, $\sigma(t)$ becomes sensitive to noise in the radial projection and is therefore evaluated from the large-scale background and treated as piecewise constant across such intervals.

Equation~\eqref{eq:Ueff_def_app_revised} follows the modified-Taylor-hypothesis logic in that it treats the dominant outward-propagating component as frozen in a frame moving at $\mathbf{V}_p^{\mathrm{Bulk}}\pm V_A^{\mathrm{Bulk}}\hat{\mathbf{b}}_0$ \citep{Klein_2015}, with the spacecraft motion included through $\mathbf{V}_{sc}(t)$.

\subsection{Parker-spiral reference: theoretical form and implemented direction}\label{app:ps_reference}

For steady radial outflow of speed $U(r)$ in the presence of solar rotation at angular frequency $\Omega$, the Parker-spiral field in heliocentric spherical coordinates $(r,\vartheta,\varphi)$ may be written as
\begin{equation}
B_{PS,r}(r)=B_{r0}\left(\frac{r_0}{r}\right)^2,\qquad
B_{PS,\vartheta}(r)=0,\qquad
B_{PS,\varphi}(r)=-\,\frac{\Omega\,r\sin\vartheta}{U(r)}\,B_{PS,r}(r),
\label{eq:PS_components}
\end{equation}
where $r_0$ is an inner reference radius and $B_{r0}$ fixes the normalization.
In this work the Parker-spiral reference enters only through its direction in the RT plane.
We adopt $\Omega = 2.9 \times 10^{-6}~\mathrm{s}^{-1}$, take $r(t)$ as the spacecraft heliocentric distance, and set $r_0 = 10\,R_{\odot}$ as the source-surface radius \citep{Bruno_1997}.
Following \citealt{Fargette2021}, we define the RT-plane angle
\begin{equation}
\alpha_{PS}(t)\equiv \arctan\!\left(\frac{\Omega\,[\,r(t)-r_0\,]}{V_{r,\mathrm{LP}}(t)}\right),
\label{eq:alphaPS_def}
\end{equation}
where $V_{r,\mathrm{LP}}(t)$ is the radial speed after applying a 2-hour low-pass filter.
To avoid sector-polarity bias in $\theta_{PS}$, we introduce a polarity factor evaluated from the large-scale background,
\begin{equation}
\sigma_{PS}(t)\equiv \mathrm{sgn}\!\big(B_0^{r}(t)\big),
\label{eq:sigmaPS_def}
\end{equation}
with the same piecewise-constant treatment as in \S~\ref{app:kinetic_windows} when $|B_0^{r}|$ is small.
The adopted Parker-spiral unit vector is then
\begin{equation}
\hat{\mathbf{b}}_{PS}(t)\equiv \sigma_{PS}(t)\,\big(\cos\alpha_{PS}(t),\,\sin\alpha_{PS}(t),\,0\big)_{\mathrm{RTN}}.
\label{eq:bhatPS_def}
\end{equation}
Because $\hat{\mathbf{b}}_{PS}$ is restricted to the RT plane, out-of-plane components contribute to $\theta_{PS}$ by construction; this is an explicit choice of reference geometry.

When a Parker-spiral vector amplitude is required (Appendix~\ref{app:wkb_theta_ps}), we set
\begin{equation}
\mathbf{B}_{PS}(t)\equiv |B(t)|\,\hat{\mathbf{b}}_{PS}(t),
\label{eq:BPS_amp_def}.
\end{equation}

\subsection{Deflection angles, background fields, and switchback identification}\label{app:def_angles}

Switchback diagnostics are defined as deflection angles between the measured magnetic-field direction $\hat{\mathbf{b}}(t)\equiv \mathbf{B}(t)/B(t)$ and a prescribed reference direction $\hat{\mathbf{b}}_{\mathrm{ref}}(t)$.
The generic deflection angle is
\begin{equation}
\theta_{\mathrm{ref}}(t)\equiv \cos^{-1}\!\Big(\hat{\mathbf{b}}(t)\cdot\hat{\mathbf{b}}_{\mathrm{ref}}(t)\Big).
\label{eq:theta_ref_def}
\end{equation}
To mitigate dependence on background definition, we retain three standard reference choices in parallel: a moving mean \citep{Bandyopadhyay_alfven}, a moving median \citep{deWit_sbs_WT}, and a Parker-spiral direction \citep[e.g.,][]{Horbury_2018, Fargette2021}. These choices yield overlapping but non-identical event populations and can introduce systematic differences in occurrence rates and statistical distributions \citep{Badman2021, Fargette2022}.

For a prescribed scale $\ell^\ast=N\,d_i$ with window width $w_{\ell^\ast}(t)$ from Eq.~\eqref{eq:w_of_lstar_def_revised}, we define the local reference fields and directions by
\begin{equation}
\mathbf{B}^{\mathrm{mea}}_{\ell^\ast}(t)\equiv \langle \mathbf{B}\rangle_{w_{\ell^\ast}}(t),
\quad
\hat{\mathbf{b}}^{\mathrm{mea}}_{\ell^\ast}(t)\equiv \frac{\mathbf{B}^{\mathrm{mea}}_{\ell^\ast}(t)}{\left|\mathbf{B}^{\mathrm{mea}}_{\ell^\ast}(t)\right|},
\qquad
\mathbf{B}^{\mathrm{med}}_{\ell^\ast}(t)\equiv \mathrm{med}_{w_{\ell^\ast}}[\mathbf{B}](t),
\quad
\hat{\mathbf{b}}^{\mathrm{med}}_{\ell^\ast}(t)\equiv \frac{\mathbf{B}^{\mathrm{med}}_{\ell^\ast}(t)}{\left|\mathbf{B}^{\mathrm{med}}_{\ell^\ast}(t)\right|}.
\label{eq:Brefs_and_bhats_lstar_def}
\end{equation}
The mean- and median-referenced deflections (used throughout the main text) are obtained by specializing Eq.~\eqref{eq:theta_ref_def}:
\begin{equation}
\theta_{\mathrm{mea}}(t;\ell^\ast)\equiv \theta_{\mathrm{ref}}(t)\Big|_{\hat{\mathbf{b}}_{\mathrm{ref}}=\hat{\mathbf{b}}^{\mathrm{mea}}_{\ell^\ast}},
\qquad
\theta_{\mathrm{med}}(t;\ell^\ast)\equiv \theta_{\mathrm{ref}}(t)\Big|_{\hat{\mathbf{b}}_{\mathrm{ref}}=\hat{\mathbf{b}}^{\mathrm{med}}_{\ell^\ast}}.
\label{eq:theta_mea_med_def}
\end{equation}
Using $\hat{\mathbf{b}}_{PS}(t)$ from Appendix~\ref{app:ps_reference}, the Parker-spiral--referenced deflection is
\begin{equation}
\theta_{PS}(t)\equiv \theta_{\mathrm{ref}}(t)\Big|_{\hat{\mathbf{b}}_{\mathrm{ref}}=\hat{\mathbf{b}}_{PS}(t)}.
\label{eq:thetaPS_def_methods}
\end{equation}
Using the large-scale median background $\hat{\mathbf{b}}_0(t)$ from Eq.~\eqref{eq:B0_def_app}, define the moving-median background deflection
\begin{equation}
\theta_{0}(t)\equiv \theta_{\mathrm{ref}}(t)\Big|_{\hat{\mathbf{b}}_{\mathrm{ref}}=\hat{\mathbf{b}}_{0}(t)}.
\label{eq:theta0_def_methods}
\end{equation}

\section{WKB expectation for Parker-spiral deflections as a function of $M_a$}\label{app:wkb_theta_ps}

This appendix is written to make the dependency chain used in the main text explicit:
(i) map the windowed Parker-spiral deflection statistic $\overline{c}_{PS}(t)=\langle\cos\Theta_{PS}\rangle_w(t)$ to a normalized Parker-spiral--referenced amplitude (Eq.~\eqref{eq:angle_amplitude_mapping});
(ii) use wave-action conservation to obtain the WKB $M_a$-dependence of a normalized amplitude (Eq.~\eqref{eq:dB_over_B_of_Ma});
(iii) define an observable proxy $\mathcal{Y}_{PS}(t)$ that combines (i)--(ii) (Eq.~\eqref{eq:YPS_def});
and (iv) test whether the conditioned mean $\langle \mathcal{Y}_{PS}\rangle_{M_a}$ follows the predicted profile $\propto g(M_a)$ (Eq.~\eqref{eq:profile_fit_def}).
All operators and background quantities are defined in Appendix~\ref{Appendix:Methods} (\S\S~\ref{app:operators_units}--\ref{app:def_angles}).

Unless stated otherwise, we evaluate the Parker-spiral diagnostics using the same window definition as the local-median background in Fig.~\ref{fig:def_angles_pdf} Panel~$(\beta)$, i.e.\ $w\equiv w_{\ell^\ast}(t)$ with $\ell^\ast=2^{14}\,d_i$.

\subsection{Parker-spiral--referenced fluctuations}\label{app:fluctuation_decomposition}

Define the Parker-spiral--referenced fluctuation
\begin{equation}
\delta\mathbf{B}_{PS}(t)\equiv \mathbf{B}(t)-\mathbf{B}_{PS}(t),
\qquad
\delta B_{PS,\mathrm{inst}}(t)\equiv \big|\delta\mathbf{B}_{PS}(t)\big|,
\label{eq:deltaBPS_inst_def}
\end{equation}
where $\mathbf{B}_{PS}(t)$ and $\theta_{PS}(t)$ are given by Eqs.~\eqref{eq:BPS_amp_def} and \eqref{eq:thetaPS_def_methods}.
For notational consistency with the main text, we also define $\Theta_{PS}(t)\equiv \theta_{PS}(t)$, so that $\overline{c}_{PS}(t)=\langle\cos\Theta_{PS}\rangle_w(t)$.
With $B(t)\equiv|\mathbf{B}(t)|$ and $B_{PS}(t)\equiv|\mathbf{B}_{PS}(t)|$, the instantaneous amplitude satisfies
\begin{equation}
\big(\delta B_{PS,\mathrm{inst}}(t)\big)^2
=
\big(B(t)-B_{PS}(t)\big)^2
+
2\,B(t)\,B_{PS}(t)\,\Big[1-\cos\Theta_{PS}(t)\Big].
\label{eq:deltaBPS_general_cos}
\end{equation}

For compactness, define
\begin{equation}
X_{PS}^2(t)\equiv 2\Big(1-\cos\Theta_{PS}(t)\Big),
\label{eq:XPS2_def}
\end{equation}
so that $0\leq X_{PS}^2\leq 4$ \citep{zhdankin_statistical_2013}. Equation~\eqref{eq:deltaBPS_general_cos} becomes
\begin{equation}
\big(\delta B_{PS,\mathrm{inst}}(t)\big)^2
=
\big(B(t)-B_{PS}(t)\big)^2
+
B(t)\,B_{PS}(t)\,X_{PS}^2(t).
\end{equation}

Using the rolling average operator $\langle\cdot\rangle_w$ defined in Eq.~\eqref{eq:window_average_def_app}, define the rms amplitude
\begin{equation}
\delta B_{PS}(t)\equiv \Big[\big\langle \delta B_{PS,\mathrm{inst}}^2\big\rangle_{w}(t)\Big]^{1/2}.
\label{eq:deltaBPS_window_def}
\end{equation}
In intervals dominated by RD-like, nearly constant-$|B|$ Alfv\'enic rotations, the field-strength mismatch term is small on the analysis window, i.e.\ $|B-B_{PS}|/B_{PS}\ll X_{PS}$, and $B_{PS}(t)$ varies slowly across $w$. Then
\begin{equation}
2\Big[1-\big\langle \cos\Theta_{PS}\big\rangle_{w}(t)\Big]
\simeq
\left(\frac{\delta B_{PS}(t)}{B_{PS}(t)}\right)^2,
\label{eq:angle_amplitude_mapping}
\end{equation}
so that $\langle\cos\Theta_{PS}\rangle_w$ provides a direct proxy for $\delta B_{PS}/B_{PS}$ in the RD-like, nearly constant-$|B|$ limit.

Systematic departures from Eq.~\eqref{eq:angle_amplitude_mapping} indicate that $(B-B_{PS})^2$ in Eq.~\eqref{eq:deltaBPS_general_cos} contributes non-negligibly on the same window. This occurs for compressive or pressure-balanced fluctuations.

\subsection{Theoretical framework: wave-action $M_a$-profile}\label{app:wkb_theta_ps:wkb}

Assume a slowly varying background (WKB/adiabatic ordering) and outward-dominated, weakly dissipative Alfv\'enic fluctuations within a prescribed band \citep{witham1965general, bretherton1968wavetrains, Heinemann_Olbert, Velli_93, 2015_Chandran_wave_action, Tenerani_EBM}.

Wave-action density is $\mathcal{S}\equiv \mathcal{E}/\omega_0$, with $\mathcal{E}$ the wave energy density and $\omega_0 \equiv \omega - \mathbf{k}\cdot\mathbf{U}$, and satisfies
\begin{equation}
\frac{\partial \mathcal{S}}{\partial t} + \nabla\cdot\!\left(\mathbf{V}_g\,\mathcal{S}\right) = 0,
\end{equation}
where $\mathbf{V}_g$ is the group velocity \citep{bretherton1968wavetrains}.
For steady propagation guided by a flux tube of cross-sectional area $A(r)$, with $\mathbf{V}_g$ directed along the tube, integration gives the flux-tube invariant
\begin{equation}
A(r)\,V_g(r)\,\frac{\mathcal{E}(r)}{\omega_0(r)}=\mathrm{const}.
\label{eq:AVgE_over_omega0_const}
\end{equation}

For outward Alfv\'enic packets, $V_g=U+V_A$ along the tube and $\omega_0=kV_A$.
Substituting into Eq.~\eqref{eq:AVgE_over_omega0_const} yields
\begin{equation}
A(r)\,\frac{U(r)+V_A(r)}{k(r)\,V_A(r)}\,\mathcal{E}_A(r)=\mathrm{const}.
\label{eq:preK_invariant}
\end{equation}
To obtain the invariant used in the analysis we further assume the band is tied to an (approximately) fixed spacecraft-frame frequency $\omega$ across $r$, so that $\omega\simeq (U+V_A)\,k$ implies $k(r)\propto [U(r)+V_A(r)]^{-1}$.
This step is used only to motivate the expected $M_a$-profile; the comparison in the main text is restricted to the profile of conditioned means of the empirical diagnostics defined below.
Then Eq.~\eqref{eq:preK_invariant} reduces to
\begin{equation}
\mathcal{K}\equiv A(r)\,\frac{\big(U(r)+V_A(r)\big)^2}{V_A(r)}\,\mathcal{E}_A(r)=\mathrm{const}.
\label{eq:K_def}
\end{equation}

In the selected band we approximate Alfv\'enic polarization with small compressive contribution and take $\mathcal{E}_A\simeq |\delta\mathbf{B}|^2/\mu_0$ (up to order-unity partition factors).
Normalizing Eq.~\eqref{eq:K_def} by the tube mass flux $\rho U A$ gives
\begin{equation}
\mathcal{V}_W^2 \equiv \frac{\mathcal{K}}{\rho U A}
= \frac{V_A\,(U+V_A)^2}{U}\left(\frac{\delta B_{\mathrm{rms}}}{B_0}\right)^2,
\label{eq:VW_def}
\end{equation}
where $\delta B_{\mathrm{rms}}^2\equiv \langle |\delta\mathbf{B}|^2\rangle$ denotes the band variance (with averaging as defined in \S~\ref{app:operators_units}).
Writing $M_a\equiv U/V_A$, Eq.~\eqref{eq:VW_def} is equivalent to
\begin{equation}
\left(\frac{\delta B_{\mathrm{rms}}}{B_0}\right)^2
=
\left(\frac{\mathcal{V}_W}{U}\right)^2 \frac{M_a^{3}}{(M_a+1)^2}.
\label{eq:dB_over_B_of_Ma}
\end{equation}
Define
\begin{equation}
g(M_a)\equiv \frac{M_a^3}{(M_a+1)^2},
\qquad
f(M_a)\equiv \frac{(M_a+1)^2}{M_a^3}.
\label{eq:g_f_def}
\end{equation}

\subsection{Diagnostics and the $g(M_a)$ test}\label{app:wkb_theta_ps:diagnostics}

Define the windowed mean cosine of the Parker-spiral deflection angle,
\begin{equation}
\overline{c}_{PS}(t)\equiv \big\langle \cos\Theta_{PS}\big\rangle_{w}(t),
\label{eq:cbarPS_def}
\end{equation}
with $\Theta_{PS}(t)\equiv \theta_{PS}(t)$.
Under the mapping \eqref{eq:angle_amplitude_mapping} and the amplitude choice $B_{PS}=B_0$ from Eq.~\eqref{eq:BPS_amp_def}, the quantity $2\,[1-\overline{c}_{PS}(t)]$ estimates $\big(\delta B_{PS}(t)/B_0(t)\big)^2$ in the RD-like, nearly constant-$|B|$ limit.

To represent $\rho U A$ in Eq.~\eqref{eq:VW_def}, we use the spherical mass-flux proxy
\begin{equation}
\dot{M}(t)\equiv 4\pi r^2(t)\,\rho^{\mathrm{Bulk}}(t)\,U^{\mathrm{Bulk}}(t),
\label{eq:mdot_def_app}
\end{equation}
where $\rho^{\mathrm{Bulk}}$ and $U^{\mathrm{Bulk}}$ are defined in \S~\ref{app:Ma_defs}. Equation~\eqref{eq:mdot_def_app} corresponds to $\rho U A$ under $A(r)\propto r^2$; any constant tube solid-angle factor is absorbed into the fitted normalization below.

Using $\overline{c}_{PS}$, define
\begin{equation}
Y_{\mathrm{LHS}}(t)\equiv 2\,\big(U^{\mathrm{Bulk}}(t)\big)^2\,\Big[1-\overline{c}_{PS}(t)\Big],
\label{eq:Ylhs_def}
\end{equation}
and its mass-flux-weighted form
\begin{equation}
\mathcal{Y}_{PS}(t)\equiv \dot{M}(t)\,Y_{\mathrm{LHS}}(t).
\label{eq:YPS_def}
\end{equation}
The role of $\mathcal{Y}_{PS}$ is to translate the geometric amplitude proxy \eqref{eq:angle_amplitude_mapping} into the wave-action normalization in Eq.~\eqref{eq:VW_def}, so that its conditioned mean can be compared to the $M_a$-profile in Eq.~\eqref{eq:dB_over_B_of_Ma}.

As an amplitude-based alternative, use the windowed RMS amplitude about the Parker-spiral background $\delta B_{PS}(t)$ defined in \S~\ref{app:fluctuation_decomposition} and define
\begin{equation}
R_{PS}(t)\equiv \frac{\delta B_{PS}(t)}{B_0(t)},
\label{eq:Rps_def}
\end{equation}
\begin{equation}
\mathcal{X}_{PS}(t)\equiv \dot{M}(t)\,\big(U^{\mathrm{Bulk}}(t)\big)^2\,R_{PS}^2(t).
\label{eq:Xps_def_app}
\end{equation}

The wave-action prediction for conditioned profiles is expressed as
\begin{equation}
\big\langle \mathcal{Y}_{PS}\big\rangle_{M_a}\simeq \alpha\,g(M_a),
\label{eq:profile_fit_def}
\end{equation}
where $\langle\cdot\rangle_{M_a}$ denotes the bin-average operator \eqref{eq:bin_average_def_app} applied to samples selected by $M_a^{\mathrm{Bulk}}(t)$ (Eq.~\eqref{eq:Ma_bulk_def_app}), and $\alpha$ absorbs the inner-boundary normalization and any constant geometric factor implicit in Eq.~\eqref{eq:mdot_def_app}.

\bibliography{sample701}{}
\bibliographystyle{aasjournalv7}



\end{CJK*}
\end{document}